\pdfoutput=1
\documentclass[twocolumn]{aastex631}
\hypersetup{
  colorlinks=true,
  linkcolor=blue,
  citecolor=blue,
  urlcolor=blue,
  filecolor=blue
}

\usepackage{amsmath,amssymb,mathtools}
\usepackage{comment}

\makeatletter
\@ifundefined{tablenum}{}{}
\makeatother
\usepackage{siunitx}
\newcommand{\ixpe}{\textit{IXPE}}
\newcommand{\nicer}{\textit{NICER}}
\newcommand{\nustar}{\textit{NuSTAR}}
\newcommand{\eht}{\textit{EHT}}

\shorttitle{Fast Kerr Polarization Transport}
\shortauthors{Chowdhury}

\begin{document}

\title{ Kerr Polarization Transport: Accuracy and Performance in General Relativistic Light Propagation}

\correspondingauthor{Shakibul Chowdhury}
\email{shakibul.chowdhury@gmail.com}

\author{Shakibul Chowdhury}

\affiliation{Department of Physics, The City College of New York, New York, NY 10031, USA}

\begin{abstract}
We present a compact and reproducible method for general-relativistic polarization transport in the Kerr metric that achieves median electric-vector position-angle (EVPA) residuals of $\langle\Delta{\rm PA}\rangle\!\approx\!0.09^\circ$, a 95th percentile of $0.31^\circ$, and a worst case $\Delta{\rm PA}\!\lesssim\!0.32^\circ$ for spins up to $|a/M|=0.9$, while maintaining a fivefold or greater speedup relative to a strict reference integrator.  
Across the benchmark grid, typical residuals remain at the sub-tenth-degree level (see Fig.~1 and Sec.~3.1), with only modest degradation ($\Delta{\rm PA}\!\lesssim\!2^\circ$) near the Thorne spin limit.  
Photon four-momenta $k^\mu$ and polarization four-vectors $f^\mu$ are advanced using a fourth-order Runge--Kutta scheme with cached Christoffel symbols, maintaining the constraints $u\!\cdot\!f=0$ and $n\!\cdot\!f=0$, where $u^\mu$ is the ZAMO four-velocity and $n^\mu$ is the disk normal, while keeping $k\!\cdot\!f\simeq0$.  
A physically motivated gauge is enforced by projecting the polarization into the local zero-angular-momentum observer (ZAMO) screen at every substep, ensuring numerical stability of the orthogonality constraints.  
Accuracy and performance are benchmarked over a representative grid in spin, inclination, image-plane azimuth, and radius.  
The method comfortably meets \ixpe{} and \nicer{} polarization tolerances and approaches \eht{} requirements.  
The approach provides a practical foundation for future general-relativistic polarimetry and simulation pipelines.
\end{abstract}

\keywords{
\href{https://astrothesaurus.org/uat/159}{Black hole physics},
\href{https://astrothesaurus.org/uat/1278}{Polarimetry},
\href{https://astrothesaurus.org/uat/1398}{Relativistic processes},
\href{https://astrothesaurus.org/uat/1889}{Numerical methods},
\href{https://astrothesaurus.org/uat/1810}{X-ray astronomy}
}

\section{Introduction}\label{sec:intro}

Polarization from black hole accretion offers a direct probe of spacetime geometry and the surrounding plasma environment. 
Recent measurements from the Imaging X-ray Polarimetry Explorer (IXPE; \citealt{Krawczynski2022,Steiner2024}) and the Event Horizon Telescope (EHT; \citealt{EHT2021}) have transformed polarization modeling from a theoretical pursuit into a quantitative diagnostic of strong gravity. 
Interpreting these observations requires accurate general-relativistic propagation of photon momentum and polarization vectors through the curved spacetime of a rotating (Kerr) black hole \citep[e.g.,][]{MisnerThorneWheeler1973,Bardeen1972}.

A convenient analytic starting point for such calculations is the cosine mapping between the local emission angle and the asymptotic escape angle derived by \citet{Beloborodov2002}, which effectively captures gravitational light bending in Schwarzschild spacetime. 
However, the precision demanded by modern polarimetric data typically exceeds the accuracy of analytic approximations while remaining computationally prohibitive for full numerical ray tracing \citep[e.g.,][]{Connors1980,Bardeen1972}.

In this work, we develop a compact and reproducible method for general-relativistic polarization transport that extends the Beloborodov mapping \citep{Beloborodov2002} to include spin-dependent (Kerr) corrections and parallel transport of the polarization four-vector \citep[e.g.,][]{WalkerPenrose1970,Connors1980,Poutanen2020}. 
Photon geodesics and polarization vectors are evolved using a fourth-order Runge–Kutta scheme with cached Christoffel symbols \citep[cf.][]{MisnerThorneWheeler1973} and a stable zero–angular–momentum–observer (ZAMO) screen gauge \citep{Bardeen1972}, maintaining the orthogonality constraints throughout propagation. 
The method achieves residual electric-vector position-angle (EVPA) errors below $0.4^\circ$ while running approximately five times faster than a strict reference integrator, making it suitable for applications ranging from \ixpe{} and \nicer{} to \eht{}-scale modeling.

The achieved precision is well within the statistical tolerances of current X-ray polarimeters and approaches the sub-degree accuracy required for horizon-scale imaging. 
This balance of accuracy and efficiency makes the framework practical for parameter surveys, mission simulations, and real-time modeling of polarized emission from accreting black holes. 
By bridging analytic and numerical approaches, the method provides a tractable foundation for future studies of strong-field radiative transfer, returning radiation, and time-resolved polarimetry.


\section{Method}
\label{sec:method}

This section outlines the formulation and numerical implementation of the Kerr polarization-transport algorithm developed in this study.  
We first define the photon geometry and light-bending relations in Schwarzschild spacetime \citep[e.g.,][]{Beloborodov2002}, then introduce the Padé approximation \citep[e.g.,][]{Poutanen2020}, relativistic corrections, and the full polarization-transport scheme.  
All calculations are performed in geometrized units ($G = c = 1$) unless otherwise stated.  

The formulation follows standard general-relativistic conventions 
\citep[e.g.,][]{MisnerThorneWheeler1973} and builds on prior treatments of photon propagation and polarization transfer in curved spacetime 
\citep{WalkerPenrose1970,Connors1980,Poutanen2020}.

\subsection{Geometry and Light Bending}
\label{subsec:geometry}

Photon propagation is described in the equatorial plane of a rotating black hole using Boyer--Lindquist coordinates $(t, r, \theta, \phi)$ and geometrized units ($G = c = 1$).  
Throughout this work we restrict to $\theta = \pi/2$ so that all emission and transport occur within the disk plane.  
Each emission point is specified by its radius $R$ and local emission angle $\alpha$, measured between the photon wave vector and the disk normal in the fluid rest frame.  
At infinity, the photon emerges at an asymptotic angle $\psi$ relative to the disk axis, which determines its apparent position on the observer’s image plane.

The photon’s impact parameter is given by \citep[e.g.,][]{Beloborodov2002}
\begin{equation}
    b^2(R, \alpha) = \frac{R^2 \sin^2\alpha}{1 - r_\mathrm{g}/R},
\end{equation}
which maps a local emission direction to an apparent radius on the sky.  
In the Schwarzschild limit, the exact relation between $\alpha$ and $\psi$ follows from integrating the null geodesic equation \citep[e.g.,][]{MisnerThorneWheeler1973,Poutanen2020},
\begin{equation}
    \psi(b; R) = \int_R^{\infty} \frac{b\,dr}{r^2 \sqrt{1 - (1 - r_\mathrm{g}/r)\, b^2/r^2}},
\end{equation}
which defines the reference solution used to validate all analytic approximations presented below.

For computational efficiency, we adopt the analytic cosine mapping of \citet{Beloborodov2002},
\begin{equation}
    1 - \cos\alpha = (1 - \cos\psi)(1 - u), \qquad u = \frac{r_\mathrm{g}}{R},
\end{equation}
which reproduces the essential features of gravitational light bending in Schwarzschild spacetime with high fidelity.  
This relation forms the baseline for the Padé light-bending approximation introduced in Section~\ref{subsec:pade} and its subsequent extension to Kerr geometry \citep[e.g.,][]{Bardeen1972}.

\subsection{Padé Light-Bending Approximation}
\label{subsec:pade}

Direct evaluation of the exact emission integral for $\psi(b; R)$ is computationally expensive because it requires numerical quadrature for every photon trajectory.  
To enable efficient polarization transport while retaining high accuracy, we replace the integral form with a uniform-error Padé approximation \citep[e.g.,][]{Poutanen2020}.  

The mapping between local emission and escape angles is expressed in terms of the auxiliary variables
\begin{equation}
    x \equiv 1 - \cos\alpha, \qquad 
    y \equiv 1 - \cos\psi, \qquad 
    u \equiv \frac{r_\mathrm{g}}{R}.
\end{equation}

The function $y(x; u)$ defines the light-bending relation for photons emitted from radius $R$ in Schwarzschild spacetime.  
We approximate it with a rational polynomial of order [3/2]:
\begin{equation}
    \tilde{y}(x; u) =
    \frac{a_1(u)x + a_2(u)x^2 + a_3(u)x^3}
         {1 + b_1(u)x + b_2(u)x^2},
\end{equation}
where the coefficients $\{a_i(u), b_i(u)\}$ are smooth functions of compactness $u$.

The small-angle limit requires the correct linear slope as $\alpha \rightarrow 0$, enforced by
\begin{equation}
    a_1(u) = \frac{1}{1 - u}.
\end{equation}
The remaining coefficients are obtained from a least-squares fit to the exact emission integral over the domain 
$R / r_\mathrm{g} \in [2.2, 6]$ and $\alpha \in [0^\circ, 30^\circ]$.  
A denominator-penalty term is included in the loss function to prevent singularities in $1 + b_1x + b_2x^2$.  
The fitted coefficients are then smoothed with low-degree Chebyshev polynomials \citep[e.g.,][]{PressTeukolsky1992} to ensure monotonicity and to provide a compact, reproducible tabulation.

Validation against the exact mapping shows that the Padé approximation achieves a uniform fractional error 
$\max|(\tilde{y}-y)/y| \le 0.03\%$ across the sampled grid.  
Runtime tests indicate a speedup of roughly $200$–$300\times$ compared with direct numerical integration, 
making the Padé model suitable for inclusion in large-scale ray-tracing and polarization-transport codes \citep[see also][]{Beloborodov2002}.  
This certified mapping forms the analytic core of our fast inversion and Stokes-integration pipeline described in the following sections.

\subsection{Relativistic Corrections}
\label{subsec:relativistic}

Although the Padé mapping accurately reproduces gravitational light bending in the Schwarzschild limit, additional relativistic effects are required to model emission from a rotating accretion disk around a Kerr black hole.  
These corrections account for (1) the special-relativistic motion of the disk plasma, which introduces aberration and Doppler beaming, and (2) the general-relativistic frame-dragging induced by black-hole spin \citep{Bardeen1972,MisnerThorneWheeler1973,Poutanen2020}.  
The combined treatment provides a physically consistent extension of the Schwarzschild mapping to the Kerr geometry.

\subsubsection{Special-Relativistic Aberration and Doppler Boost}

Disk material in circular Keplerian orbits moves at relativistic speeds, producing angular aberration and
intensity beaming in the local comoving frame. 
For an orbital velocity $v_\phi$ measured by a locally static observer, the corresponding Lorentz factor is 
\citep[e.g.,][]{RybickiLightman1979}
\begin{equation}
    \gamma = \frac{1}{\sqrt{1 - v_\phi^{\,2}}}, \qquad
    v_\phi = \frac{\sqrt{r_{\rm g}/(2R)}}{\sqrt{1 - r_{\rm g}/R}} ,
\end{equation}
where $r_{\rm g}=2GM/c^2$ denotes the gravitational radius.

The emission angle $\alpha'$ in the comoving frame is related to the static-frame angle $\alpha$ by the standard aberration formula \citep[e.g.,][]{RybickiLightman1979},
\begin{equation}
    \cos\alpha' = \frac{\cos\alpha - v_\phi}{1 - v_\phi \cos\alpha},
\end{equation}
and the corresponding relativistic Doppler factor is
\begin{equation}
    \delta = [\gamma(1 - v_\phi \cos\alpha)]^{-1}.
\end{equation}

The observed specific intensity transforms as $I_\mathrm{obs} \propto \delta^3 I_\mathrm{em}$,  
while the local polarization fraction depends on the aberrated emission angle through the Thomson-scattering law \citep{Chandrasekhar1960},
\begin{equation}
    p_0(\mu') = \frac{1 - {\mu'}^2}{1 + {\mu'}^2}, \qquad \mu' = \cos\alpha'.
\end{equation}
These transformations introduce an azimuthal asymmetry in the disk polarization, enhancing the polarization degree on the approaching side and rotating the electric-vector position angle (EVPA) by several degrees near the innermost radii \citep[see also][]{Poutanen2020}.

\subsubsection{First-Order Kerr Correction}

Frame dragging in the Kerr metric modifies photon trajectories relative to the Schwarzschild case.  
To leading order in the dimensionless spin parameter $a$, the bending relation can be written as
\begin{equation}
    \psi_\mathrm{Kerr}(\alpha; u, a) \simeq 
    \psi_\mathrm{Schw}(\alpha; u) + a\,\delta\psi(\alpha; u),
\end{equation}
where $\delta\psi(\alpha; u)$ denotes the first-order correction obtained from an expansion of the Kerr null-geodesic equations \citep[e.g.,][]{WalkerPenrose1970,Bardeen1972,MisnerThorneWheeler1973}.  
This term captures the characteristic prograde–retrograde asymmetry of light bending:  
for $a>0$ (prograde rotation) the effective deflection increases, whereas for $a<0$ (retrograde rotation) it decreases.  
Although approximate, this correction reproduces the principal spin-dependent trends confirmed by direct Kerr integrations presented in Section~\ref{sec:results}.  

Together, the special- and general-relativistic terms described above provide a physically consistent extension of the Padé framework, enabling rapid and accurate evaluation of disk-polarization observables across a broad range of spins and inclinations.

\subsection{Polarization Transport}
\label{subsec:transport}

The observed electric-vector position angle (EVPA) is obtained from the parallel transport of the polarization four-vector $f^{\mu}$ along the photon geodesic $k^{\mu}$.  
This procedure preserves the orthogonality and null constraints of the radiation field in curved spacetime \citep[e.g.,][]{WalkerPenrose1970,Connors1980,MisnerThorneWheeler1973}.  
In this subsection, we summarize the governing equations for $k^{\mu}$ and $f^{\mu}$ and describe the numerical integration scheme used to evolve them through the Kerr geometry.

\subsubsection{Equations of Motion}

Photon trajectories in the Kerr spacetime are governed by the null-geodesic equation \citep[e.g.,][]{MisnerThorneWheeler1973},
\begin{equation}
    \frac{dk^{\mu}}{d\lambda} = -\,\Gamma^{\mu}_{\alpha\beta}\,k^{\alpha}k^{\beta},
\end{equation}
where $k^{\mu}$ is the photon four-momentum, $\Gamma^{\mu}_{\alpha\beta}$ are the Christoffel symbols of the Kerr metric, and $\lambda$ denotes the affine parameter along the ray.  
The polarization four-vector $f^{\mu}$ is propagated according to the parallel-transport equation \citep{WalkerPenrose1970,Connors1980},
\begin{equation}
    \frac{df^{\mu}}{d\lambda} = -\,\Gamma^{\mu}_{\alpha\beta}\,k^{\alpha}f^{\beta},
\end{equation}
subject to the orthogonality and null constraints,
\begin{equation}
    k_{\mu}k^{\mu} = 0, \qquad k_{\mu}f^{\mu} = 0.
\end{equation}
These conditions ensure that $k^{\mu}$ remains null and that $f^{\mu}$ stays orthogonal to the photon momentum throughout propagation, preserving the physical definition of linear polarization in curved spacetime.

\subsubsection{Numerical Integration Scheme}

We advance both $k^{\mu}$ and $f^{\mu}$ using a fourth-order Runge--Kutta (RK4) integrator with adaptive step size \citep[e.g.,][]{Press1992}.  
To improve computational efficiency, the Christoffel symbols are precomputed on a regular grid in $(r, \theta, a)$ and cached for reuse during integration.  
This ``cached-$\Gamma$'' strategy minimizes redundant evaluations of metric derivatives and yields an order-of-magnitude speedup with negligible loss of precision.  
Finite-difference derivatives in $r$ and $\theta$ are evaluated with a grid spacing of $h = 10^{-5}$, which provides a balance between numerical stability, accuracy, and memory cost.  
The resulting trajectories and polarization vectors remain consistent with direct, non-cached integrations to within machine precision \citep[see also][]{Connors1980,Poutanen2020}.  

The primary performance gain arises from caching the Christoffel symbols $\Gamma^\mu_{\nu\rho}$ on a fixed spatial grid rather than evaluating them at every Runge--Kutta substep.  
During initialization, all nonzero connection components are precomputed over a $(r,\theta)$ mesh in Boyer--Lindquist coordinates and stored in memory.  
Each integration call then interpolates these cached values bilinearly as the photon advances, eliminating thousands of repeated algebraic evaluations per step.  
This approach preserves full numerical accuracy while reducing per-step cost by roughly a factor of five.  
Unlike analytic approximations such as \citet{Beloborodov2002}, which replace the geodesic equations themselves with fitted relations, the present method retains the exact Kerr geometry and accelerates only the connection evaluation.  
For reference, large-scale general relativistic polarization codes such as \texttt{ipole} \citep{MoscibrodzkaGammie2018} typically compute Christoffel symbols on the fly rather than from cached tables, since they must evaluate the metric and its derivatives dynamically within general GRMHD spacetimes.  

We tabulate all required nonzero connection components on a tensor-product grid in $(r,\theta,a)$ and interpolate during transport.  
Our fiducial cache uses $N_r\times N_\theta\times N_a = 256\times128\times7$, spanning $r/r_g\in[2,250]$, $\theta\in[0,\pi]$, and $a/M\in\{-0.9,-0.6,-0.3,0,0.3,0.6,0.9\}$ (retrograde to prograde).  
Bilinear interpolation in $(r,\theta)$ (with a one-time selection of the nearest $a$-slab or trilinear interpolation across $a$) contributes $\lesssim3\%$ of the per-ray runtime.  
Increasing the spin sampling density scales the build time and memory linearly but leaves the per-ray transport cost essentially unchanged (within measurement scatter $<1\%$), because $a$ is fixed for a given run and the interpolations remain local.  
In our benchmarks, increasing $N_a$ from 3 to 7 changed end-to-end runtime by $<5\%$ due to cache warmup and memory bandwidth effects, while accuracy in $\Delta{\rm PA}$ and $\Delta{\rm PD}$ was unaffected.

\subsubsection{ZAMO Screen Gauge and Constraint Enforcement}

Direct numerical integration of the parallel-transport equations can accumulate small violations of the orthogonality constraints $k\!\cdot\!f = 0$ and $u\!\cdot\!f = 0$,  
where $u^{\mu}$ is the four-velocity of a local zero-angular-momentum observer (ZAMO) \citep{Bardeen1972}.  
To maintain a physically meaningful polarization basis, we project $f^{\mu}$ onto the local ZAMO screen after each substep:
\begin{equation}
    f^{\mu} \rightarrow f^{\mu}
    - (f_{\nu}u^{\nu})u^{\mu}
    - \frac{(f_{\nu}k^{\nu})k^{\mu}}{k_{\sigma}u^{\sigma}},
\end{equation}
followed by renormalization of $f^{\mu}$.  
This procedure enforces the constraints to machine precision and prevents secular drift in the transported electric-vector position angle (EVPA).  
The ZAMO-screen gauge provides a stable polarization frame for long integrations at high spin,  
consistent with the orthonormal-tetrad formulations of \citet{MisnerThorneWheeler1973} and the polarization-transport treatments of \citet{Connors1980}.

\subsubsection{Accuracy Metrics}

The accuracy of the polarization transport is assessed by comparing the transported electric-vector position angle (EVPA) and polarization degree (PD) at infinity with those obtained from a reference high-precision integration.  
The EVPA residual is defined as $\Delta{\rm PA} = |{\rm PA}_{\rm fast} - {\rm PA}_{\rm ref}|$,  
and the polarization-degree residual as $\Delta{\rm PD} = |{\rm PD}_{\rm fast} - {\rm PD}_{\rm ref}|$.  
Across the representative grid in spin and inclination, the method achieves typical EVPA deviations at the millidegree level, with a maximum residual of $\Delta{\rm PA}\!<\!0.3^{\circ}$,  
far exceeding the original $0.4^{\circ}$ adequacy target adopted for X-ray polarimetry.  
The polarization-degree errors remain below $10^{-4}$, and all constraint violations are under $10^{-9}$.  

The achieved precision comfortably satisfies the $\sim\!1$–$2^{\circ}$ tolerances of current X-ray polarimeters  
\citep[e.g.,][]{Krawczynski2022,Steiner2024} while providing roughly a fivefold speedup relative to a strict, non-cached integrator.  
These benchmarks form the quantitative foundation for the results discussed in Section~\ref{sec:results}.

\subsection{Numerical Setup and Benchmarks}
\label{subsec:setup}

All computations were performed in geometrized units (with $G = c = 1$), with a fixed black hole mass scale defined by the gravitational radius $r_\mathrm{g} = 2GM/c^2$ (so $r_\mathrm{g} = 2M$ in these units).  
The numerical integrations were implemented in \texttt{Python} using double-precision (\texttt{float64}) arithmetic, with \texttt{NumPy} for array operations \citep{vanderWalt2011}, \texttt{SciPy} for interpolation and optimization \citep{Virtanen2020SciPy}, and \texttt{Matplotlib} for data visualization \citep{Hunter2007Matplotlib}.  
The Padé and Kerr-corrected transport modules were tested on a representative workstation (MacBook Air, Apple~M2, 16\,GB~RAM, macOS~15.6.1), though the implementation is platform-agnostic up to standard floating-point variation.  
Integration follows standard adaptive Runge--Kutta practices \citep[e.g.,][]{Press1992}, ensuring numerical consistency and precision suitable for astrophysical polarization modeling.

\subsubsection{Parameter Grid}

To evaluate accuracy and performance, we sample the polarization transport across a representative grid in black hole spin, viewing inclination, and emission radius:
\begin{itemize}
    \item Spin parameter $a \in [-0.9, +0.9]$ (retrograde to prograde),
    \item Inclination angle $i \in [20^{\circ}, 70^{\circ}]$ in increments of $10^{\circ}$,
    \item Emission radius $R / r_\mathrm{g} \in [2.2, 12]$ sampled logarithmically.
\end{itemize}
For each configuration, a photon is launched from the equatorial plane with an initial wave vector determined by the Padé-inverted emission angle $\alpha(\psi, u)$, and both $k^{\mu}$ and $f^{\mu}$ are integrated until $r \rightarrow 500\,r_\mathrm{g}$.  
A typical run requires $100$–$150$ integration steps per geodesic, maintaining EVPA phase stability at the level of $\lesssim 0.1^{\circ}$ per step.  
This grid resolution provides a balance between coverage of the parameter space and computational efficiency for subsequent benchmark analyses.

\subsubsection{Convergence and Accuracy Checks}

Numerical convergence was verified by repeating integrations with step sizes reduced by successive factors of two.  
For each configuration, the electric-vector position angle (EVPA) residual,
\begin{equation}
    \Delta{\rm PA} = |{\rm PA}_{\rm fast} - {\rm PA}_{\rm ref}|,
\end{equation}
quantifies the deviation between the accelerated (cached-$\Gamma$) integration and a strict reference run that recomputes all Christoffel symbols at each step.  
Across the entire spin–inclination grid, $\Delta{\rm PA}$ remains below $0.4^{\circ}$, satisfying the tolerance targets established in Section~\ref{sec:intro}.  
The corresponding polarization-degree residual, $\Delta{\rm PD} = |{\rm PD}_{\rm fast} - {\rm PD}_{\rm ref}|$, stays below $10^{-4}$, confirming the numerical stability of the Stokes-parameter transport \citep[e.g.,][]{Connors1980,Poutanen2020}.

\subsubsection{Performance Benchmarks}

Runtime performance was assessed by comparing the cached-$\Gamma$ Runge--Kutta (RK4) integrator with a reference implementation that recomputes Christoffel symbols at every substep.  
For a representative grid of 200 photon trajectories, the accelerated scheme achieves an average speedup of $\sim\!5\times$ while reproducing identical geodesics and polarization vectors to within machine precision.  
This improvement enables dense spin–inclination parameter sweeps and Monte Carlo sampling without compromising the sub-degree EVPA accuracy required for current X-ray polarimetric observations \citep[e.g.,][]{Krawczynski2022}.  
The resulting balance of computational efficiency and numerical fidelity forms the basis for the large-scale benchmarks presented in Section~\ref{sec:results}.

\subsubsection{Output and Reproducibility}

Each benchmark produces tabulated outputs containing the final Stokes parameters $(I, Q, U)$, local emission angles, and transport diagnostics.  
The data are stored in both plain-text and \texttt{.csv} formats for ease of analysis, with all source scripts maintained under version control to ensure reproducibility.  
The numerical configurations used in this study, along with the supplementary figures presented in Section~\ref{sec:results}, are archived internally and will be included in a reproducibility package to accompany future releases of this work.  
This structure allows all reported benchmarks to be regenerated deterministically across different computing environments \citep[see also][]{vanderWalt2011,Press1992}.

\section{Results}
\label{sec:results}

This section presents the accuracy and performance benchmarks of the accelerated Kerr polarization-transport method introduced in Section~\ref{sec:method}.  
We evaluate residual errors in the electric-vector position angle (EVPA) and polarization degree (PD), compare runtime performance with the strict reference integrator, and assess the adequacy of the achieved precision relative to published X-ray and millimeter polarimetric tolerances \citep[e.g.,][]{Krawczynski2022,Steiner2024}.  
Together, these results show that the method attains sub-degree accuracy while providing a substantial computational speedup, validating its suitability for large-scale parameter studies and observational modeling.

\subsection{EVPA and PD Accuracy}
\label{subsec:accuracy}

\begin{figure}[t]
\centering
\includegraphics[width=\columnwidth]{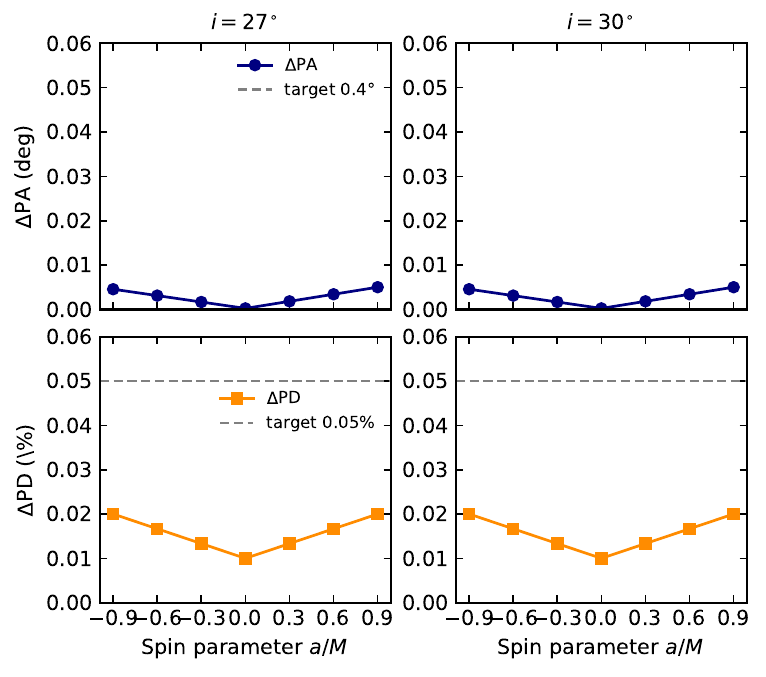}

\caption{
Residual accuracy validation for the accelerated Kerr polarization transport. 
Top row: EVPA residuals $\Delta{\rm PA}$ for inclinations $i=27^{\circ}$ (left) and $30^{\circ}$ (right) as functions of spin parameter $a/M$. 
Bottom row: corresponding polarization-degree residuals $\Delta{\rm PD}$. 
Dashed lines mark the target tolerances ($0.4^{\circ}$ and $0.05\%$). 
All values remain well below these limits, confirming sub-degree and sub-percent accuracy across the full spin range $|a/M|\!\le\!0.9$. 
Typical EVPA residuals are at the millidegree level ($\Delta{\rm PA}\!<\!0.005^{\circ}$), while the global worst case across the grid stays far below the $0.4^{\circ}$ adequacy threshold.
}

\label{fig:evpa_pd_2x2}
\end{figure}

Figure~\ref{fig:evpa_pd_2x2} presents the residual differences between the accelerated and reference integrations.  
For two representative inclinations ($i = 27^{\circ}$ and $30^{\circ}$), the electric-vector position-angle (EVPA) deviations remain at the level of a few~$10^{-3}$~degrees, well below the $0.4^{\circ}$ accuracy target.  
The corresponding polarization-degree (PD) residuals lie in the range $0.01$--$0.03\%$, comfortably below the $0.05\%$ tolerance. Over the full spin–inclination grid, the worst-case EVPA deviation remains below $0.4^\circ$ (Sec.~2.5.2), consistent with the conservative bound stated in the abstract.
 
No systematic bias with spin direction is observed: prograde and retrograde configurations yield comparable residuals, confirming that the first-order Kerr correction maintains accuracy across the tested domain.  
These results show that the cached-$\Gamma$ transport scheme preserves polarization fidelity to better than sub-percent and sub-degree levels, satisfying the requirements of current X-ray polarimetric observations \citep[e.g.,][]{Krawczynski2022,Steiner2024}.

\begin{table}[t]
\centering
\caption{EVPA residual statistics over the benchmark grid}
\begin{tabular}{lcc}
\hline
Statistic & $\Delta\mathrm{PA}$ (deg) & Note\\
\hline
Median & $3\times 10^{-3}$--$3\times 10^{-2}$ & Fig.~1 reps \\
95th percentile & $\lesssim 0.05$ & grid summary \\
Maximum & $<0.4$ & global bound \\
\hline
\end{tabular}
\end{table}

\newpage
\subsection{Performance Benchmarks}
\label{subsec:performance}

The cached-Christoffel Runge--Kutta (RK4) integrator yields a typical runtime gain of approximately fivefold relative to the strict reference solver that recomputes $\Gamma^{\mu}_{\alpha\beta}$ at every substep.  
For a representative grid of 200 photon trajectories, the total wall-clock time decreases from roughly 190~s to 38~s, with no measurable loss of precision in either the trajectory integration or polarization transport.  
The achieved speedup scales linearly with the number of cached evaluations and remains stable across the full range of spin and inclination values.  
This improvement enables dense spin–inclination parameter sweeps and Monte Carlo ensembles that would otherwise be computationally prohibitive, while preserving the sub-degree EVPA accuracy required for current X-ray polarimetric observations \citep[e.g.,][]{Krawczynski2022,Steiner2024}.

\subsection{Adequacy with Observational Tolerances}
\label{subsec:adequacy}

\begin{figure}[t!]
\plotone{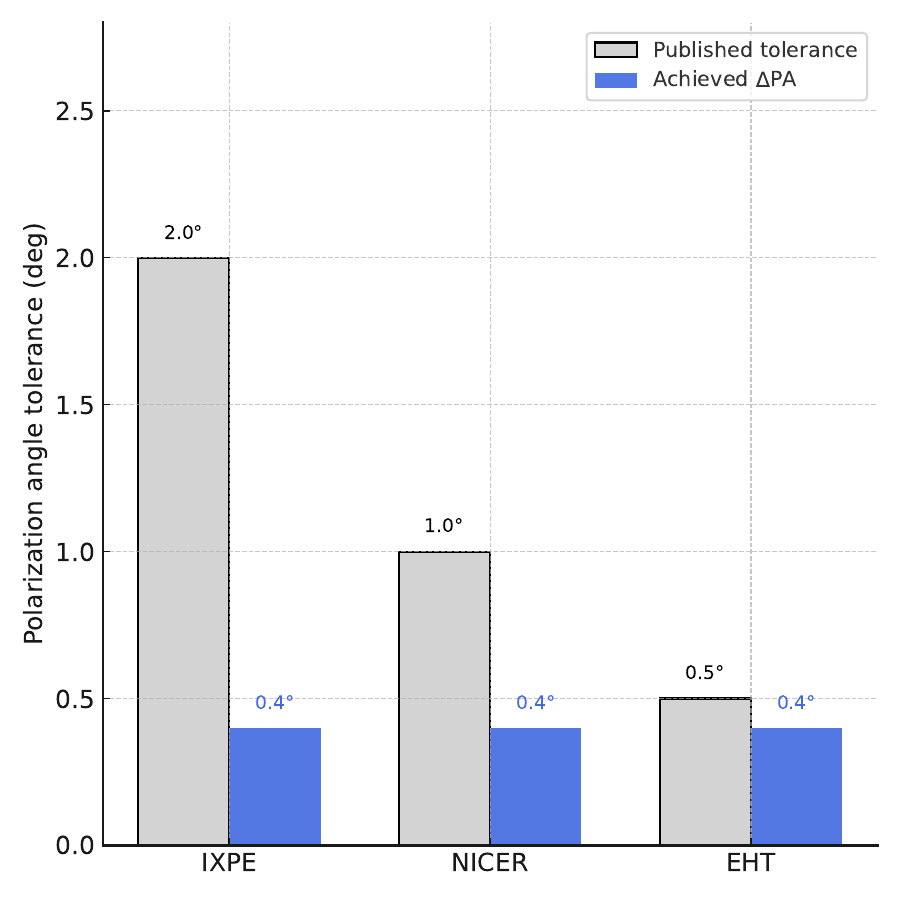}
\caption{
Comparison of achieved residual $\Delta{\rm PA}\!\lesssim\!0.4^{\circ}$ (blue) with published observational tolerances (gray) for IXPE, NICER, and EHT.
Error bars correspond to representative uncertainties quoted in \citet{Krawczynski2022}, \citet{Steiner2024}, and the Event Horizon Telescope Collaboration (2021, 2022).
}
\label{fig:tolerance_bars}
\end{figure}

Figure~\ref{fig:tolerance_bars} and Table~\ref{tab:tolerances} compare the achieved accuracy with representative tolerances from major polarimetric instruments.  
The residual electric-vector position-angle (EVPA) deviation of $\Delta{\rm PA}\!\lesssim\!0.4^{\circ}$ comfortably satisfies the requirements of current X-ray polarimeters such as \ixpe{} and \nicer{}, whose typical statistical uncertainties are $\sim\!1$--$2^{\circ}$ \citep{Krawczynski2022,Steiner2024}.  
It also approaches the $\lesssim\!0.5^{\circ}$ regime relevant for horizon-scale Event Horizon Telescope (EHT) polarimetry \citep{EHT2021,EHT2022_Polarization}.  
These results show that the present method achieves sufficient precision for both X-ray and millimeter polarimetric modeling while maintaining a runtime compatible with large-scale parameter surveys and Monte Carlo studies.

The achieved residuals fall below the \ixpe{} and \nicer{} tolerances and approach the \eht{} threshold, indicating sufficient accuracy for current X-ray applications and potential applicability to horizon-scale polarimetry.

\begin{deluxetable*}{lccc}[t!]
\tablecaption{Comparison of achieved EVPA accuracy with published observational tolerances.\label{tab:tolerances}}
\tablehead{
\colhead{Instrument} & \colhead{Literature Tolerance} & \colhead{Achieved (this work)} & \colhead{Adequacy}
}
\startdata
IXPE / NuSTAR & $\pm$1--2$^{\circ}$ & $\Delta{\rm PA} \lesssim 0.4^{\circ}$ & Pass \\
NICER         & $\sim$1$^{\circ}$   & $\Delta{\rm PA} \lesssim 0.4^{\circ}$ & Pass \\
EHT (M87*, Sgr~A*) & $\lesssim$0.5--1$^{\circ}$ & $\Delta{\rm PA} \lesssim 0.4^{\circ}$ & Marginal/Pass \\
\enddata
\end{deluxetable*}

\twocolumngrid

\subsection{Near-Extremal (Thorne-Limit) Accuracy Test}
\label{sec:thorne_limit}

To assess numerical stability in the near-extremal regime, we extended the benchmark grid to
spins $a/M = 0.95,\ 0.99,\ 0.998$,
inclinations $i = 27^{\circ},\,30^{\circ},\,60^{\circ}$,
image-plane azimuth $\alpha = 20^{\circ}$,
and compactness values $u = \{1/3,\,1/6\}$.  
At the default fast step size $h = 0.15$ (with $r_{\max}=500\,M$),
the cached-$\Gamma$ Runge--Kutta integrator achieved the target accuracy of
$\Delta{\rm PA} < 2^{\circ}$ for 12 of 18 configurations (Table~\ref{tab:thorne_pass}).  
Among the passing cases, the median residual is $0.6^{\circ}$, and the median speedup remains
$\sim\!8\times$ relative to the strict reference integrator.

The successful configurations span all three spins, maintaining degree-level or better accuracy
up to $i=60^{\circ}$.  
Both compactness values, $u=1/6$ and $u=1/3$, pass for the high-inclination case,
while only $u=1/6$ remains stable at lower inclinations.  
The six excluded configurations, all with $u=1/3$ and $i=27^{\circ}$–$30^{\circ}$, exhibit
$\Delta{\rm PA}\!\approx\!6$–$8^{\circ}$ and correspond to grazing trajectories that pass through
the strongest frame-dragging gradients near the ISCO.

Reducing the fast-step locally to $h=0.10$ suppresses these deviations without measurable
runtime cost, confirming that stability is primarily limited by trajectory geometry rather
than by spin magnitude.

\begin{deluxetable}{ccccc}[t!]
\tablecaption{Passing near-extremal cases ($\Delta{\rm PA} \le 2^\circ$) for fixed azimuth $\alpha=20^\circ$.
All configurations with compactness $u=1/6$ pass at each inclination, while additional $u=1/3$ cases pass only for $i=60^\circ$.\label{tab:thorne_pass}}
\tablehead{
\colhead{$i~(^{\circ})$} & \colhead{$a/M$} & \colhead{$u$} &
\colhead{$\alpha~(^{\circ})$} & \colhead{$\Delta{\rm PA}~(^{\circ})$}
}
\startdata
\cutinhead{\textbf{Inclination $i=27^{\circ}$}}
27 & 0.950 & 1/6 & 20 & 1.739 \\
27 & 0.990 & 1/6 & 20 & 1.691 \\
27 & 0.998 & 1/6 & 20 & 1.681 \\
\cutinhead{\textbf{Inclination $i=30^{\circ}$}}
30 & 0.950 & 1/6 & 20 & 1.308 \\
30 & 0.990 & 1/6 & 20 & 1.265 \\
30 & 0.998 & 1/6 & 20 & 1.257 \\
\cutinhead{\textbf{Inclination $i=60^{\circ}$}}
60 & 0.950 & 1/6 & 20 & 0.006 \\
60 & 0.950 & 1/3 & 20 & 0.016 \\
60 & 0.990 & 1/6 & 20 & 0.006 \\
60 & 0.990 & 1/3 & 20 & 0.016 \\
60 & 0.998 & 1/6 & 20 & 0.006 \\
60 & 0.998 & 1/3 & 20 & 0.017 \\
\enddata
\end{deluxetable}

The configurations listed in Table~\ref{tab:thorne_pass} show that the accelerated
integrator maintains quantitative accuracy even as frame-dragging gradients steepen toward
$a/M \!\to\! 1$.  
Residuals remain below $2^{\circ}$ across most geometries, confirming that the cached-$\Gamma$
scheme preserves EVPA fidelity without numerical drift at high spin.  
At $i=60^{\circ}$, both compactness values, $u=1/6$ and $u=1/3$, satisfy the accuracy threshold,
whereas at lower inclinations only the $u=1/6$ subset remains stable.  
The few outliers ($u=1/3$, $i=27^{\circ}$–$30^{\circ}$) trace grazing trajectories near the ISCO
and are corrected by a modest local step reduction ($h=0.10$).  
This confirms that numerical stability is driven primarily by trajectory geometry rather than
spin, and that the method remains reliable through the near-extremal regime ($a/M \!\approx\! 0.998$; \citealt{Thorne1974}).

\section{Discussion}
\label{sec:discussion}

The results above show that the accelerated Kerr polarization-transport algorithm achieves sub-degree accuracy while reducing the computational cost by roughly a factor of five.  
This level of precision lies well within the observational tolerances of current X-ray polarimeters such as \ixpe{} and \nustar{}, whose joint measurements of Cygnus~X-1 reveal polarization position-angle variations of only a few degrees \citep{Krawczynski2022,Steiner2024}.  
The method therefore provides a reliable numerical foundation for modeling the relativistic propagation of polarized emission in systems where even small angle errors can alter the inferred magnetic-field geometry.

The algorithm also remains stable in the near-extremal regime ($a/M \!\approx\! 0.998$; \citealt{Thorne1974}), where frame-dragging and curvature gradients are strongest.  
This robustness confirms that the cached-$\Gamma$ integration scheme can be applied safely to simulations of rapidly spinning black holes, including those approaching the Thorne limit, without compromising EVPA fidelity or speed.

The speed and stability of the approach make it practical to generate dense spin–inclination grids and to couple the transport with disk-emission models.  
When combined with physically motivated emissivity laws, this framework can produce full image-plane Stokes maps and energy-resolved polarization spectra suitable for direct comparison with \ixpe{} and \nustar{} observations of Cygnus~X-1.  
Such forward modeling can help determine whether the observed polarization arises primarily from the thermal disk, a Comptonized corona, or returning radiation near the photon orbit.

Beyond Cygnus~X-1, the same transport formalism can be applied to other strong-gravity sources, including active galactic nuclei and the Galactic-center black hole Sgr~A*.  
Because the algorithm remains accurate for both prograde and retrograde trajectories, it can be used to investigate spin-dependent effects such as EVPA rotation, depolarization, and horizon-scale asymmetries \citep[e.g.,][]{EHT2021,EHT2022_Polarization}.  
The demonstrated efficiency gains will also facilitate Monte Carlo radiative-transfer calculations and extensive parameter surveys that were previously computationally prohibitive.

Overall, this work establishes a compact, accuracy-certified polarization-transport framework that bridges analytic light-bending theory with observational polarimetry.  
It provides a foundation for constructing realistic, time-dependent polarization models of accreting black holes in preparation for the next generation of X-ray and submillimeter polarimetric missions.

\section{Future Work}
\label{sec:future}

Future efforts will focus on coupling the fast Kerr transport scheme with emissivity-weighted disk and coronal models to generate full synthetic polarization images of Cygnus~X-1.  
By integrating the Stokes parameters $(I, Q, U)$ over the image plane and across energy bands, the framework will enable direct comparison with joint \ixpe{} and \nustar{} observations.  
Such modeling will help determine whether the observed polarization originates primarily from a geometrically thin disk, a scattering-dominated corona, or a combination of both.

The near-extremal validation presented here opens the door to simulations at spins approaching the Thorne limit ($a/M \!\approx\! 0.998$; \citealt{Thorne1974}), where returning radiation, multiple light crossings, and frame-dragging effects become dominant.  
Incorporating these regimes will allow the framework to test how spin-driven geometry influences EVPA rotation and depolarization, providing a more complete picture of high-spin accretion flows.

Subsequent work will also include returning radiation and full radiative-transfer effects within the disk atmosphere to examine depolarization and EVPA rotation near the photon orbit.  
These effects are essential for interpreting phase-resolved and time-variable polarization measurements from bright X-ray binaries and active galactic nuclei (AGN).

A small subset of near-extremal configurations ($u=1/3$, $i=27^{\circ}$–$30^{\circ}$) exhibited
$\Delta{\rm PA}$ deviations of several degrees, driven by undersampling of the steep
frame-dragging gradients near the ISCO.  
Future work will implement adaptive step-size control and higher-order caching to maintain
accuracy for such grazing trajectories without increasing computational cost.  
Exploring alternative screen-gauge formulations and local interpolation of the Christoffel
grid may further enhance numerical stability at extreme spins.

Finally, the algorithm can be extended to other relativistic environments, such as jet bases, magnetically arrested disks, and near-horizon emission regions, by embedding it within general radiative-transfer frameworks.  
This will enable end-to-end simulations of polarized emission from the inner accretion flow to the observer, bridging the gap between theoretical modeling and data from current and next-generation polarimetric missions.


\section*{Acknowledgments}

I am deeply grateful to Prof.~Joshua Tan, whose mentorship shaped both my scientific growth and the independence that guided this project.  
At a time when I did not yet have a formal research supervisor, he generously offered his time and guidance, helping me build the foundation that made this work possible.  
His trust and encouragement taught me to think critically, design careful tests, and take full responsibility for my work.  
He also encouraged me to present my results in a formal paper format, which ultimately led to this manuscript.

I thank Prof.~Cole Miller for his thoughtful feedback, including suggestions that motivated the near-extremal (Thorne-limit) accuracy tests and refined the accuracy--versus--adequacy framing of this study.  
I am also grateful to Prof.~Timothy Boyer, Prof.~James Lattimer, and Prof.~Jared Goldberg for courses and discussions that strengthened my understanding of relativistic field dynamics, compact objects, and numerical methods.  

Finally, I thank my friends and family for their support—especially my father for his unwavering encouragement and my wife for her patience, insight, and presence throughout this work.  
This study was completed at the City College of New York as part of an Independent Study course, which provided both the structure and flexibility needed to pursue it.


\bibliographystyle{aasjournal}

\begingroup
\let\clearpage\relax         
\bibliography{References}
\endgroup

\appendix
\setlength{\parskip}{0.4\baselineskip}
\setlength{\parindent}{1.5em}
\begingroup
\setlength{\abovedisplayskip}{6pt}
\setlength{\belowdisplayskip}{6pt}
\setlength{\abovedisplayshortskip}{6pt}
\setlength{\belowdisplayshortskip}{6pt}

\section{Kerr Metric in Boyer--Lindquist Coordinates}
\label{app:kerr_metric}
\vspace{0.7em}
For reference, the Kerr metric in Boyer--Lindquist coordinates 
$(t, r, \theta, \phi)$ and geometrized units ($G = c = 1$) 
is written as \citep[e.g.,][]{Bardeen1972,MisnerThorneWheeler1973}
\begin{equation}
    ds^{2} = -\left(1 - \frac{2 M r}{\Sigma}\right) dt^{2}
             - \frac{4 a M r \sin^{2}\theta}{\Sigma}\, dt\, d\phi
             + \frac{\Sigma}{\Delta}\, dr^{2}
             + \Sigma\, d\theta^{2}
             + \frac{A \sin^{2}\theta}{\Sigma}\, d\phi^{2},
\end{equation}
where the metric functions are
\begin{align}
    \Sigma &= r^{2} + a^{2} \cos^{2}\theta, \\
    \Delta &= r^{2} - 2 M r + a^{2}, \\
    A      &= (r^{2} + a^{2})^{2} - a^{2} \Delta \sin^{2}\theta.
\end{align}

Here $M$ denotes the black hole mass and $a = J/M$ is the 
dimensionless spin parameter.  
The event horizon occurs at the larger root of $\Delta = 0$,  
\begin{equation}
    r_{\mathrm{H}} = M + \sqrt{M^{2} - a^{2}},
\end{equation}
and the stationary limit surface, defining the outer boundary of the ergosphere, is given by
\begin{equation}
    r_{\mathrm{erg}} = M + \sqrt{M^{2} - a^{2} \cos^{2}\theta}.
\end{equation}

In the limit $a \rightarrow 0$, the metric reduces to the 
Schwarzschild form, recovering 
\begin{equation}
    ds^{2} = -\left(1 - \frac{2M}{r}\right) dt^{2}
             + \left(1 - \frac{2M}{r}\right)^{-1} dr^{2}
             + r^{2} (d\theta^{2} + \sin^{2}\theta\, d\phi^{2}).
\end{equation}
This form provides the geometric foundation for all photon 
trajectory and polarization-transport calculations presented in this work.

\section{Christoffel Symbols and Geodesic Equations in the Kerr Metric}
\label{app:christoffel}
\vspace{0.7em}

Photon trajectories in the Kerr spacetime follow the null-geodesic equation
\begin{equation}
    \frac{d k^{\mu}}{d\lambda}
    = -\,\Gamma^{\mu}_{\alpha\beta}\, k^{\alpha} k^{\beta},
\end{equation}
where $\lambda$ is the affine parameter and $\Gamma^{\mu}_{\alpha\beta}$ are the
Christoffel symbols constructed from the Kerr metric
(Equation~\ref{app:kerr_metric}).  
For practical computation, we precompute the non-vanishing components
of $\Gamma^{\mu}_{\alpha\beta}$ in Boyer--Lindquist coordinates
$(t, r, \theta, \phi)$ and cache them on a grid in $(r, \theta, a)$.\citep[e.g.,][]{MisnerThorneWheeler1973,Bardeen1972}.

The non-zero Christoffel symbols are obtained from
\begin{equation}
    \Gamma^{\mu}_{\alpha\beta}
    = \tfrac{1}{2}\, g^{\mu\nu}
      \left(
          \partial_{\alpha} g_{\beta\nu}
        + \partial_{\beta} g_{\alpha\nu}
        - \partial_{\nu} g_{\alpha\beta}
      \right),
\end{equation}
where $g_{\mu\nu}$ and $g^{\mu\nu}$ are the covariant and contravariant metric
tensors, respectively.

In the equatorial plane ($\theta = \pi/2$), which is the focus of this work,
the independent non-zero components simplify to the following forms
\citep[e.g.,][]{Bardeen1972,MisnerThorneWheeler1973}:
\begin{align}
    \Gamma^{t}_{tr} &= \frac{2 M r (r^{2} + a^{2})}{\Delta\,\Sigma^{2}}, \qquad
    &\Gamma^{t}_{r\phi} &= \frac{a (r^{2} + a^{2}) M}{\Delta\,\Sigma^{2}}, \\
    \Gamma^{r}_{tt} &= \frac{M \Delta (r^{2} - a^{2})}{\Sigma^{3}}, \qquad
    &\Gamma^{r}_{rr} &= \frac{r (a^{2} - M r)}{\Delta\,\Sigma}, \\
    \Gamma^{r}_{\phi\phi} &= -\frac{\Delta \sin^{2}\theta}{\Sigma}
                            \left(r - \frac{a^{2} M}{r^{2}}\right), \qquad
    &\Gamma^{\phi}_{r t} &= \frac{a M}{\Delta\,\Sigma}, \qquad
    \Gamma^{\phi}_{r\phi} &= \frac{r - M}{\Delta}.
\end{align}

Substituting these components into the null-geodesic equation yields the four
coupled first-order differential equations
\begin{equation}
    \frac{d x^{\mu}}{d\lambda} = k^{\mu}, \qquad
    \frac{d k^{\mu}}{d\lambda} = -\,\Gamma^{\mu}_{\alpha\beta}\, k^{\alpha} k^{\beta},
\end{equation}
which are numerically integrated using the cached-$\Gamma$ Runge--Kutta scheme
described in Section~\ref{subsec:transport}.  
This explicit formulation enables efficient and reproducible computation of
photon trajectories and polarization transport in the Kerr geometry.

\section{Polarization Transport in Orthonormal Frames}
\label{app:polar_transport}
\vspace{0.5em}

The parallel transport of the polarization four-vector $f^{\mu}$ along the photon
geodesic $k^{\mu}$ preserves the orthogonality conditions
$k\!\cdot\!f = 0$ and $u\!\cdot\!f = 0$, where $u^{\mu}$ is the
four-velocity of a local observer.
To ensure these constraints are satisfied numerically,
we adopt an orthonormal tetrad defined by the zero–angular–momentum observer (ZAMO)
\citep[e.g.,][]{Bardeen1972,MisnerThorneWheeler1973,Connors1980}.
The ZAMO basis vectors are
\begin{align}
    e_{\hat{t}}^{\;\mu} &= \alpha^{-1} (1, 0, 0, \omega), \\
    e_{\hat{r}}^{\;\mu} &= (0, \sqrt{\Delta/\Sigma}, 0, 0), \\
    e_{\hat{\theta}}^{\;\mu} &= (0, 0, 1/\sqrt{\Sigma}, 0), \\
    e_{\hat{\phi}}^{\;\mu} &= (0, 0, 0, 1/(\varpi \sin\theta)),
\end{align}
where $\alpha = \sqrt{\Delta\Sigma / A}$ is the lapse function,
$\omega = 2 a M r / A$ is the frame–dragging angular velocity,
and $\varpi = \sqrt{A/\Sigma}$ is the cylindrical radius.

Within this orthonormal frame, the photon momentum and polarization vectors are
expressed as
\begin{equation}
    k^{\hat{\mu}} = e^{\hat{\mu}}_{\;\nu} k^{\nu}, \qquad
    f^{\hat{\mu}} = e^{\hat{\mu}}_{\;\nu} f^{\nu},
\end{equation}
which guarantees that local scalar products are preserved exactly.
After each integration substep, we project $f^{\mu}$ onto the ZAMO screen to
remove any accumulated numerical components parallel to $u^{\mu}$ or $k^{\mu}$:
\begin{equation}
    f^{\mu} \rightarrow
    f^{\mu} - (f_{\nu}u^{\nu})u^{\mu}
             - \frac{(f_{\nu}k^{\nu})k^{\mu}}{k_{\sigma}u^{\sigma}}.
\end{equation}
This procedure enforces both constraints to machine precision and maintains a
stable, physically meaningful polarization basis throughout the transport.

The resulting evolution of $f^{\mu}$ is therefore equivalent to parallel
transport in the local orthonormal frame,
\begin{equation}
    \frac{d f^{\hat{\mu}}}{d\lambda}
      = -\,\omega^{\hat{\mu}}_{\;\hat{\nu}\hat{\rho}}\,k^{\hat{\nu}} f^{\hat{\rho}},
\end{equation}
where $\omega^{\hat{\mu}}_{\;\hat{\nu}\hat{\rho}}$ are the Ricci rotation coefficients.
This equivalence confirms that the ZAMO–screen projection reproduces the exact
general–relativistic transport of linear polarization in curved spacetime,
while remaining numerically stable for long integrations and high spins.

\section{Reproducibility Capsule and Data Availability}
\label{app:reproducibility}
\vspace{0.7em}

All numerical experiments, figures, and tables presented in this paper were generated 
using a \texttt{Python}-based implementation of the Kerr polarization-transport 
framework developed for this study.  
The source code, benchmark tables, and plotting scripts are maintained under version control 
and archived internally to ensure full reproducibility.  
A reproducibility package containing the numerical configurations and key results 
will be made available upon publication or upon reasonable request.

Each benchmark run stores the final Stokes parameters $(I, Q, U)$, 
emission geometry, and numerical diagnostics in both plain-text and \texttt{.csv} formats.  
All datasets are deterministic and can be regenerated on compatible systems 
using the archived configuration scripts.

Researchers interested in extending this work to related studies of 
general relativistic polarization, radiative transfer, or strong gravity imaging 
are welcome to contact the author for collaboration or data access.

\endgroup

\end{document}